\documentclass[copyright,creativecommons]{eptcs}
\providecommand{\event}{CCA 2010} 
\usepackage{breakurl}             
\usepackage{amssymb}
\usepackage{amsthm}

\theoremstyle{definition}
\newtheorem{defn}{Definition}[section]

\newtheorem{assump}[defn]{Assumption}

\theoremstyle{plain}
\newtheorem{thm}[defn]{Theorem}
\newtheorem{lem}[defn]{Lemma}
\newtheorem{cor}[defn]{Corollary}
{\bf}{\it}
\def\proof{\noindent{\bf Proof.\ }}
\def\endpf{\hfill$\Box$\par\bigskip}

\newcommand\set[1]{\{#1\}}
\def\Nset{{\mathbb{N}}}

\def\Qset{{\mathbb{Q}}}
\def\Rset{{\mathbb{R}}}

\def\AA{{\mathcal A}}
\def\MM{{\mathcal M}}
\def\RR{{\mathcal R}}

\def\s{{\Sigma^*}}
\def\om{{\Sigma^\omega }}
\def\tm{{\widetilde{\mu}}}
\def\dmu{\delta_\mu}
\def\dtmu{\delta_{\tm}}

\newcommand{\an}{\ \mbox{and}\ }
\newcommand{\mto}{\rightrightarrows}

\def \In{\subseteq}
\def \dom{{\rm dom}}

\newcommand\tuple[1]{\langle#1\rangle}

\title{Complete Multi-Representations of Sets in a Computable Measure Space}
\author{Yongcheng Wu  \thanks{The author has been partially supported by DFG (Deutsche Forschungsgemeinschaft)     and the Scientific Fund of NUIST(S8107319001).}
 \institute{College of Mathematics and Physics\\ Nanjing University of Information Science and Technology\\ 210044 Nanjing,  China}
 \email{macswu@163.com}}
\def\titlerunning{Complete Multi-Representations}
\def\authorrunning{Yongcheng Wu}

\begin{document}
\maketitle
\begin{abstract}  In a recent paper, two  multi-representations for the measurable sets in a computable measure space have been introduced, which prove to be topologically complete w.r.t. certain topological properties. In this contribution, we show them recursively complete w.r.t. computability of  measure and set-theoretical operations.
\end{abstract}

\section{Introduction} In computable analysis, computability concepts depend critically on  representations of computational objects. Different representations of a same set of objects can be compared under two kinds of reductions: continuous reductions $\leq_t$ and  computable reductions $\leq$, which are string functions transform names under one naming system to names under  another one. Most interesting are the complete (multi-)representations among a naturally arising class of naming systems.
\begin{defn}\label{d12}
 Let $\Phi$ be a class of  naming systems of a set $X$. A naming
system $\delta\in\Phi$ is said {\em topologically/(recursively) complete} in $\Phi$, iff $\phi\leq_t\delta$ resp.
 $\phi\leq\delta$ for any $\phi\in\Phi$.
  \end{defn}

 For instance, recall that a representation $\delta$ is said {\em admissible } w.r.t. a topology $\tau$ if it is topologically complete among all the continuous representations w.r.t. $\tau$ and the Cantor topology on strings. Such admissible representations play important role in the topological approach to computable analysis.

  Computability frameworks of Lebesgue measure and integration have been addressed by different schools in computable analysis.
    Ker-I Ko\cite{Ko91} used oracle Turing machines to represent real sets and functions and studied  polynomial time complexity of them.
 Weihrauch\cite{Wei99} investigated computability of measures and integration on the unit interval in the type-2 theory of effectivity.
 Edalat\cite{Eda95, Eda09} constructed a domain theoretical framework for Lebesgue measures and integrals.
  Wu and Weihrauch\cite{WW06} showed how to construct a measure from an abstract Stone integration.  Wu and Ding\cite{WD05, WD06}  considered computability of measure and set-theoretical operations in the more general situation of a computable measure space as introduced by \cite{WW06}.

   Recently, the author suggests  another pair of  multi-representations, $\dmu$ and $\dtmu$, for the measurable sets in a computable measure space. They have been proven to be  topologically complete in a certain sense, see \cite{Wu10}. In this paper, we will explore computability of measure and set-theoretical operations w.r.t. them. The results show that $\dmu$ entails stronger computability than any of the multi-representations applied in \cite{WD05,WD06}.  Such results give rise to the recursive completeness of $\dmu$. Then we will discuss computability of set operations w.r.t. $\dtmu$ and show as a corollary the recursive completeness of $\dtmu$.  Proofs of the results are omitted in this extended abstract, which will be given in a separate  paper.

\section{Preliminaries}

\subsection{Limit relations}
   A {\em limit (convergence) relation}, say $\to_X$, on
a non-empty set $X$, is a relation  appointing
 points in $X$ to sequences $(x_n)$ in $X$, i.e. $\to_X\In
 X^{\omega}\times
 X$. If $(x_n)\to_X x$, we say  that $(x_n)$ \emph{converges} to $x$, where $(x_n)$ is called a  {\em $\to_X$-convergent sequence} and  $x$ is called a {\em limit} of
 $(x_n)$.

A pair $( X, \to_X)$ will be called a {\em limit space} if and
only if the limit relation $\to_X$ on $X$ satisfies the
following three axioms (cf. \cite{Hyl79,MS02}):
\begin{enumerate}
 \item[(L1)]\quad  $(x)\to_X x$;
 \item[(L2)]\quad If $(x_n)\to_X x$ then $(x_{n_k})_k\to_X x$, where $(x_{n_k})_k$ is a subsequence of $(x_n)$;
 \item[(L3)]\quad If $(x_n)$ is a sequence such that any subsequence of $(x_n)$ has a subsequence converging to $x$, then
 $(x_n)$ converges to $x$.
  \end{enumerate}
Let $Y$ be a subset of $X$. We say that $Y$ is {\em dense} in the limit space $(X,\to_X)$, if and only if
for every $x\in X$ there exists a sequence $(y_n)$ in $Y$ such that $(y_n)\to_X x$. Limit relations induce a natural notion of continuity: a function $f:\In X\to Y$ is said to be {\em continuous} w.r.t. limit relations $\to_X$ and $\to_Y$ iff $f$ preserves convergent sequences (cf. \cite{Bir36, Dud64}). Sometimes this notion of continuity is called {\em sequentially continuity} to differ with that defined in terms of topologies.

\subsection{Computable analysis}
We brief here the type-2 theory of effectivity, TTE for short, which is a representation-based approach to  computable analysis.
Let $\Sigma$ be a finite alphabet with $\set{0,1}\In\Sigma$. Let $\s$, $\om$ be the set of finite resp. infinite strings over $\Sigma$. On $\s$ we consider the discrete topology $\tau_*$ and on $\om$ the Cantor topology $\tau_C$ generated by the basis $\set{w\om|w\in\s}$.  In the following content, assume $W, V, W_i, V_i\in\set{\s, \om}$ for all $i\in\Nset=\set{0,1,2,\ldots}$. Our computational model is a Turing machine with a one-way output tape. As allowing no revisions on its output it is suitable for computing on infinite strings of symbols. For distinction, we call it a {\em type-2 machine}, TTM for short.  A partial string function $f:\In W_1\times W_2\times\cdots\times W_n\to W_0$ is called {\em computable} iff it is computed by a TTM. Of course, if the function value is an infinite string, the machine will write step by step each symbol of it and hence never halt. We agree with such an axiom, called  the {\em finiteness property} of TTMs: {\em Each finite portion of the output is already determined by a
finite portion of the input.}  This leads to the well-known result found firstly by Grzegorczyk\cite{Grz55}: Computable string functions are continuous, as  formulated by \cite[Theorem~2.2.3]{Wei00} in terms of TTE.

 Computations  on abstract objects are realized by a TTM via naming systems. A naming system for a nonempty set $X$ is a surjective multi-function $\delta:\In W\mto X$, which is called a {\em notation} if $W=\s$ or a {\em (multi-)representation} if $W=\om$. For any $w\in \dom(\delta)$, $w$ will be called a {\em $\delta$-name} of
 $x\in X$ if and when $x\in\nu(w)$.

 \begin{defn}[continuity and computability induced by naming systems]
 Let $\delta,\gamma$ be naming systems of sets $X$ and $Y$ respectively.
  \begin{enumerate}
  \item An element $x\in X$ is called {\em $\delta$-computable} iff $x$ has a computable $\delta$-name. 
  \item A subset $Z\In X$ is called {\em $\delta$-open/r.e./decidable} iff  $\delta^{-1}[Z]$ is open/r.e./decidable. 
 \item A {\em $(\delta,\gamma)$-realization} of a multi-function $f:\In X\mto Y$ is a (single-valued) string function $F$ such that $f(x)\cap\gamma\circ F(u)\neq\emptyset $
for any $\delta$-name $u$ of $x\in\dom(f)$. 
\item  In the above case, $f$ is called {\em $(\delta,\gamma)$-computable(-continuous)} iff  $F$ is computable(continuous). 
 \item The above definitions can be extended  to Cartesian productions and multi-variable multi-functions in a natural way.
  \end{enumerate}
\end{defn}
The above definition generalizes the corresponding notions in \cite[Definition~3.1.3]{Wei00}.

\begin{defn}[reducibility between naming systems]
Let $\delta,\gamma$ be two naming systems of a set $X$.
\begin{enumerate}
  \item $\delta$ is {\em topologically/recursively reducible} to $\gamma$, written $\delta\leq_t\gamma$ resp. $\delta\leq\gamma$ iff the identity on $X$ is $(\delta,\gamma)$-continuous/computable. (Equivalent to  \cite[Definition~24.2]{Wei08})
\item $\delta<_t\gamma$ denotes that $\delta\leq_t\gamma$ and $\gamma\nleq_t\delta$. The meaning of $\delta<\gamma$ is defined accordingly.
\item  $\equiv_t$ and $\equiv$ denote the  equivalences induced by $\leq_t$ and $\leq$ respectively.
\end{enumerate}\end{defn}

We take the so-called {\em tupling
function} $\tuple{\cdot }$ to encode a finite or infinite sequence of strings as one string (cf. \cite[Definition~2.1.7]{Wei00}). For  $w\in\s$ and $q\in W$, $w\sqsubseteq
 q$ denotes that $w$ is a prefix of $q$ and  $w\lhd q$ means $\tuple{w}$ is a substring of $q$.

 We will work with  the following standard notations
$\nu_\Nset$ of $\Nset$, $\nu_\Qset$ of $\Qset$ and standard representations $\rho$ of $\Rset$, $\overline{\rho}$ and
$\overline{\rho}_>$ of $\overline{\Rset}:=\Rset\cup\set{-\infty,\infty}$ as defined in \cite{Wei00}.  A $\rho$-name encodes essentially a converging sequence of
rational intervals to represent the limit real. Concretely, $\rho\tuple{w_1,v_2,w_2,v_2,\ldots}=x$ iff $([\nu_\Qset(w_i),\nu_\Qset(v_i)])$ is a converging sequence of intervals with a unique limit point $x$. $\rho_C$ is the {\em Cauchy} representation of $\Rset$ which uses a fast Cauchy sequence of rational numbers to represent a real. Concretely,  $\rho_C\tuple{w_1,w_2,\ldots}=x$ iff $|\nu_\Qset(w_i)-x|\leq2^{-i}$
for any $i\geq1$. In this case the rational sequence will be called a  {\em
$\rho_C$-expansion} of  $x$. It is known that  $\rho\equiv\rho_{_C}\equiv\overline{\rho}|^{\Rset}$, where $\overline{\rho}|^{\Rset}$ denotes $\overline{\rho}$ restricted to the range $\Rset$.

 \subsection{Measure theory} Let $\Omega$ be a non-empty set. A
\textit{ring} on $\Omega$ is a collection of subsets of $\Omega$
closed under the formation of finite unions and differences. An
{\em algebra} (or {\em field}) on $\Omega$ is a ring on $\Omega$
that contains $\Omega$.  A $\sigma$\textit{-algebra} (or $\sigma$-{\em field}) on $\Omega$ is an algebra  on $\Omega$
which is closed under  countable unions. Each set in a
$\sigma$-algebra is called a {\em measurable set}. For any class
$\mathcal{C}$ of sets, the minimal $\sigma$-algebra
 including $\mathcal{C}$ is called the
$\sigma$-algebra \textit{generated} by $\mathcal{C}$, written $\sigma(\mathcal{C})$.

 Suppose $\AA$ is a $\sigma$-algebra on
$\Omega$. A \textit{measure} on $\AA$ is an extended real function
$\mu:\AA\rightarrow[0,\infty]$ which is countably additive. In this case, the triple
$(\Omega,\AA,\mu)$ is  called a \textit{measure space}.

$A\bigtriangleup B:=(A\setminus B)\cup(B\setminus A)$ is the {\em
symmetric difference} of sets $A$ and $B$.  We write $A=^*B$ for
$\mu(A\bigtriangleup B)=0$ and $A\In^*B$ for $\mu(A-B)=0$.
We shall use the following terminologies:
\begin{enumerate}
 \item
$\AA_*:=\set{A\in\AA: \mu(A)<\infty}$. \item $A_{\infty}:=\AA^c_*$.
\item $\AA_{\infty\infty}:=\set{A\in\AA:\mu(A)=\mu(A^c)=\infty}$.
\item $\AA_{\infty*}:=\AA_\infty-\AA_{\infty\infty}$.
\item $A_n\nearrow A$ denotes that $(A_n)$ is an
increasing sequence of sets with $\lim_nA_n=A$.
 \item $A_n\searrow A$ denotes that
$(A_n)$ is a  decreasing sequence of sets with $\lim_nA_n=A$.
\end{enumerate}



  \section{Computable measure space}       Let $\Sigma$ be a finite alphabet. $\s$ resp. $\om$ denotes the class of finite resp. infinite strings over $\Sigma$. We take the discrete topology $\tau_*$ on $\s$ and the Cantor topology $\tau_C$ on $\om$ (Definition~2.2.2\cite{Wei00}).

\begin{defn}\cite{WW06}\label{d8}
A {\em computable measure space} is a quintuple ${\MM}=(\Omega,
\AA, \mu, \RR,\alpha)$ such that
\begin{enumerate}
\item  $(\Omega, \AA, \mu)$ is a measure space, \item
 $\RR$ is a countable ring such that $\AA=\sigma(\RR)$,
\item  $\alpha:\subseteq\s\to\RR$ is a notation of $\RR$ with recursive
domain, \item
 $\mu$ is $(\alpha,\rho)$-computable,
 \item  $(A,B)\mapsto A\cup B$ and $(A,B)\mapsto A-
B$ are computable w.r.t. $\alpha$.
\end{enumerate}
\end{defn}

Therefore, a computable measure space  $(\Omega,\AA,\mu,\RR,\alpha)$ is an abstract measure space $(\Omega,\AA,\mu)$ associated with an information structure $(\RR,\alpha)$, where $\RR$ is a countable ring generating the $\sigma$-algebra $\AA$ and $\alpha$ is a notation of $\RR$ w.r.t. which the measure $\mu$ and set-theoretical operations restricted to $\RR$ are computable.

\begin{lem}\label{l2}
In the  computable measure space
$(\Omega,\AA,\mu,\RR,\alpha)$ with $\mu(\Omega)=\infty$, there exists a computable
approximate sequence $(C_n)$  and  a computable partition
sequence $(D_n)$ of $\Omega$   so that
 \begin{enumerate}\item the multi-function $E\mapsto n$ with $E\In C_n$ is $(\alpha,\nu_\Nset)$-computable,
  \item $\mu(D_n)\geq
2^n$ and $D_n=C_n- C_{n-1}$ with $C_0:=\emptyset$.
 \end{enumerate}
\end{lem}
\proof  Firstly, let us construct effectively the sequence $(C_n)$ from the elements in $\RR$.  Since $\dom(\alpha)$ is recursive, let $(w_n)$ be a
recursive enumeration of $\dom(\alpha)$.  Let $k$ be the minimal
number such that $\mu(\bigcup_{i\leq k}\alpha(w_i))\geq 2$. Denote $C_1:=\bigcup_{i\leq k}\alpha(w_i)$. Suppose for some $n$ that $C_i$ is
defined for every $ i\leq n$. Since $\Omega=\bigcup_i \alpha(w_i)$
and $\mu(\Omega)=\infty$, there exists a minimal number $m$ such
that $\mu(\bigcup_{i\leq m}\alpha(w_i)- C_n)\geq 2^{n+1}$. Let
$C_{n+1}:=\bigcup_{i\leq m}\alpha(w_i)$. So the sequence $(C_n)$ is  recursively constructed so that, for every $n\geq 1$,
 \begin{equation}\label{e6} C_n\In C_{n+1},\ \mu(C_{n+1}-C_{n})\geq 2^{n+1} \an  \Omega=\bigcup_n C_n.\end{equation}
 Therefore, $C_n\nearrow\Omega$. Furthermore,  since the measure, union and difference  are computable w.r.t.  $\alpha$ by Definition~\ref{d8},  an $\alpha$-name of
$C_n$ can be computed for each $n\geq 1$ and therefore $(C_n)$ is
$(\nu_\mathbb{N},\alpha)$-computable.
Let us show claim (1). Given an $\alpha$-name $w$ of some set $E$, a number $n$ can be found s.t. $w_n=w$ in the recursive sequence $(w_n)$ as postulated above. This together with (\ref{e6}) implies  $E\In C_n$. Thus claim (1) holds.
As for claim (2), it suffices to set $C_0:=\emptyset$ and $D_n:=C_n- C_{n-1}$ for all $n\geq 1$.
\endpf

\begin{assump}\label{a1}
For the remaining content,  let $(\Omega,\AA,\mu,\RR,\alpha)$ be a computable measure space with $\mu(\Omega)=\infty$ and  $(C_n), (D_n)$ denote
 respectively the computable sequences as specified in Lemma~\ref{l2}.
\end{assump}

\section{Completeness of the multi-representation $\dmu$}
\begin{defn}\cite{Wu10}
The limit relation $(A_n)\to_\mu A$ for
any sequence $(A_n)$ and set $A$ in $\AA$ is defined by that
\begin{enumerate}
 \item $\lim_n\mu(A_n-A)=0$,
 \item $\lim_{n}\mu(A\cap B- A_{n})=0$ for any $B\in\AA_*$.
\end{enumerate}
\end{defn}

\begin{defn}\cite{Wu10}\label{d6}
 The multi-representation $\dmu:\In\om\mto\AA$ is
defined by that\\ $A\in\dmu\tuple{w_1,w_2,\ldots}$ iff the sequence $(A_n)$ with
$A_n:=\alpha(w_n)$ satisfies the following conditions:
 \begin{enumerate}
 \item $(A_n)\to_\mu A$,
 \item $\forall n<m$,  $\mu(A_n\bigtriangleup A_m)\leq2^{-n}$ whenever $A\in\AA_*$,
 \item$\forall n<m$, $\mu(A_n-A_m)\leq2^{-n}$,  $\mu(A_m\cap C_n- A_n)\leq2^{-n}$ and $\mu(A_n)\geq2^n$ whenever $A\in\AA_\infty$.
 \end{enumerate}
  In this case, the sequence $(A_n)$ in $\RR$ is called  a {\em $\dmu$-expansion} of  $A$.
\end{defn}
We see that $\dmu$ uses two different kinds of converging sequences under $\to_\mu$ to represent respectively the finite measurable sets $\AA_*$ and the infinitely measurable sets $\AA_\infty$.

By Lemma~\ref{l2}, the computable  sequence $(C_n)$ is indeed  a $\dmu$-expansion of $\Omega$.  Since $(E)$ is a $\dmu$-expansion of any $E\in\RR$, it follows that $\alpha\leq \dmu$.

\begin{thm}{\em\cite{Wu10}}\label{t1} If $(A_n)$ is a $\dmu$-expansion of $A$, then $A=^*\liminf_nA_n=^*\limsup_nA_n$, where $=^*$ denotes equality almost anywhere.
\end{thm}

Let  $\to_{\tau_C}$ denote the limit relation induced by the Cantor topology $\tau$ on the infinite strings $\om$.
$\dmu$ has the following topological completeness:
\begin{thm}{\em\cite{Wu10}} $\dmu$ is topologically complete in the class of  $(\to_{\tau_C},\to_\mu)$-continuous multi-representations $\phi$ of $\AA$ such that
 $\AA_\infty$ is $\phi$-open.
\end{thm}

The following lemma guarantees that $\dmu$ can differentiates effectively
$\AA_*$ from $\AA_\infty$, but cannot differentiate furthermore
$\AA_{\infty*}$ from $\AA_{\infty\infty}$ even in the topological
sense.

\begin{lem}\label{p7}\hfill
\begin{enumerate}\item $\Omega$ is $\delta_\mu$-computable, i.e. there exists a
computable $\delta_\mu$-name of $\Omega$. \item $\AA_*$ and
$\AA_\infty$ are decidable w.r.t. $\delta_\mu$.
 \item
Both $\AA_{\infty*}$ and  $\AA_{\infty\infty}$ are unopen and thus undecidable w.r.t. $\dmu$.
\end{enumerate}
\end{lem}


In \cite{WWD09}, we have shown that it is impossible for any multi-representation $\psi$ of $\AA$ to make the measure and set-theoretical operations computable on whole $\AA$. The following theorem is  nearly a reformulation of \cite[Theorem~4.4]{WWD09}.

\begin{thm}
\label{t5} Let $\psi:\In \om\to\AA$ be a
multi-representation  such that the measure $\mu$ is
$(\psi,  \overline{\rho}_{>})$-continuous. Then for
any multi-representations $\gamma,\delta$ of $\AA$, the following
statements hold:
\begin{enumerate}\item Intersection $\cap $  restricted to $\AA_{\infty\infty}$ is not $(\gamma,\delta,\psi)$-continuous. \item
Difference $-$ restricted to $\AA_{\infty\infty}$ is not
$(\gamma,\delta,\psi)$-continuous.
\item Union $\cup$ and complement $(\ )^c$  cannot  be continuous w.r.t. $\psi$ simultaneously.\end{enumerate}
The above statements hold accordingly while
``continuous" replaced by ``computable".
\end{thm}

Our studies on computability as well as incomputability of the set operations are included as one theorem:

\begin{thm}\label{t9}\hfil
\begin{enumerate}
\item $\AA_\infty$ is $\delta_\mu$-decidable.
\item
 The measure $\mu$ is
$(\delta_\mu,\overline{\rho})$-computable.
\item  Union $\cup$ is computable w.r.t. $\dmu$.
\item Intersection is $\delta_\mu$-computable on
 $\set{(A,B):A\in\AA_*,  \mbox{ or }B\in\AA_*, \mbox{ or }  A\cap B\in\AA_\infty}$,
  but not on its complement  $\set{(A,B):A,B\in\AA_{\infty\infty} \an A\cap B\in\AA_*}$.
\item Difference  is $\delta_\mu$-computable on $\set{(A,B):A\in\AA_*,
\mbox{ or }A-B\in\AA_\infty}$,
  but not on its complement  $\set{(A,B):A\in\AA_\infty,
\mbox{ and }A-B\in\AA_*}$.
 \item Complement is $\delta_\mu$-computable on
$\AA_*\cup\AA_{\infty\infty}$, but not on  $\AA_{\infty*}$.
\end{enumerate}
\end{thm}
 The above theorem shows that  $\dmu$ entails computability of set operations beyond the domains not being falsified by  the more or less general negative results as stated in Theorem~\ref{t5}.
  As a corollary, we have the following completeness theorem of $\dmu$:

\begin{cor}\label{nec3}
 $\delta_\mu$ is complete in the class $\Phi_1(\AA)$ consisting
 of all naming systems $\phi$ of  $\AA$ such that
 \begin{enumerate}
\item $\alpha\leq\phi$,
 \item  $\AA_\infty$ is $\phi$-decidable,
\item $\mu$ is
$(\phi,\overline{\rho})$-computable,
 \item Intersection $\cap$ is $\phi$-computable  on $\AA_*\times\AA$,
 \item Symmetric difference $\bigtriangleup$ is $\phi$-computable on $\AA_*\times\AA_*$.
\end{enumerate}
\end{cor}

This theorem shows that
$\dmu$-names encode exactly   sufficient and necessary
information to entail  the desired computability of the measure and set-theoretical operations.

\section{Completeness of the multi-representation   $\dtmu$}
The measure $\mu$ induces the following
 probability measure $\tm$:
\begin{equation}\label{e8}
\widetilde{\mu}(A):=\sum_{n=1}^{\infty}\frac{\mu(A\cap
D_n)}{\mu(D_n)}2^{-n}\quad (\forall A\in\AA)\end{equation}
 where $(D_n)$ is the computable partition sequence of $\Omega$ as assumed in Assumption~\ref{a1}.

\begin{defn}\label{d11}\cite{Wu10}
 The limit relation
$\to_\tm\In\AA^\omega\times\AA$  is defined by that, for
any sequence $(A_n)$ and set $A$ in $\AA$, $(A_n)\to_\tm A$ iff
 $\lim_n\tm(A_n\bigtriangleup A)=0$.
 \end{defn}

\begin{defn}\label{d10}\cite{Wu10}
The multi-representation $\dtmu:\In\om\mto\AA$ is defined by
 that $A\in\dtmu\tuple{w_1,w_2,\ldots}$ iff, for $A_n:=\alpha(w_n)$, $(A_n)\to_\tm A$ and $\tm(A_n\bigtriangleup A_m)\leq2^{-n}$ for any $n<m$.
 In this case, the sequence $(A_n)$ on $\RR$ is called  a {\em $\dtmu$-expansion} of  $A$.
 \end{defn}

$\dtmu$ is admissible w.r.t. $\to_\tm$, namely
\begin{thm}{\em \cite{Wu10}} $\dtmu$ is topologically complete among the $(\to_{\tau_C},\to_\tm)$-continuous multi-representations of $\AA$.
\end{thm}

\begin{thm}\label{t13} $\dmu<\dtmu$, i.e.
$\dmu$ is properly reducible to $\dtmu$.
\end{thm}

$\dtmu$ entails the following computability:
\begin{thm}\label{t14}\hfil
\begin{enumerate}
\item $\tm$ is $(\dtmu, \rho)$-computable.
\item  $\mu$ is $(\dtmu, \overline{\rho}_<)$-computable.
\item $(A,n)\mapsto\mu(A\cap C_n)$ is $(\dtmu, \nu_\Nset, \rho)$-computable.
\item Each set-theoretical operation is  computable w.r.t. $\dtmu$.
 \end{enumerate}
\end{thm}

As a corollary, we have
\begin{cor}\label{nec1}
$\dtmu$ is complete in the class $\Phi_2(\AA)$ consisting
 of all naming systems $\phi$ of  $\AA$ such that
\begin{enumerate}
 \item $\alpha\leq\phi$,
\item $\tm$ is $(\phi, \rho)$-computable,
\item each set-theoretical operation is computable w.r.t. $\phi$.
 \end{enumerate}
\end{cor}
By this corollary, one can see that the equivalence class of $\dtmu$ does not depend on the
computable sequence $(D_n)$ employed in the definition of $\tm$.

\bigskip
\bigskip
\noindent{\bf Acknowledgement.} The author wishes to thank the anonymous referees for their valuable questions and suggestions.

\bibliographystyle{eptcs} 

\end{document}